\documentclass[aps, reprint, superscriptaddress, nofootinbib]{revtex4-1}
\pdfoutput=1
\usepackage{amsmath, latexsym, amssymb, hyperref, graphicx, color, slashed}
\definecolor{nicered}{rgb}{0.7,0.1,0.1}
\definecolor{nicegreen}{rgb}{0.1,0.5,0.1}
\hypersetup{colorlinks, citecolor=nicegreen,linkcolor=nicered}
\begin{document}

\title{Disentangling Seesaw in the Minimal Left-Right Symmetric Model}

\author{Goran Senjanovi\'{c} and Vladimir Tello \\ \small{\it{International Centre for Theoretical Physics, Trieste, Italy}} }

\date{\today}

\begin{abstract}

    In a recent Letter we presented a systematic way of testing the seesaw origin of neutrino mass in the context of the Minimal Left-Right Symmetric Model. The essence of the program is to exploit lepton number violating decays of doubly charged scalars, particles which lie at the heart of the Higgs-mechanism-based seesaw, to probe the Dirac neutrino mass term which in turn enters directly into a number of physical processes including the decays of right-handed neutrinos into the $W$ boson and left-handed charged leptons.  
    In this longer version 
     we discuss at length these and related processes, and offer some missing technical details. We also carefully analyze the physically appealing possibility of parity conserving Yukawa sector showing that the  neutrino Dirac mass matrix can be analytically expressed as a function of light and heavy neutrino masses and mixing, without resorting to any additional discrete symmetries, 
      a context in which the seesaw mechanism can be disentangled completely.
    When parity does get broken, we show that, in the general case, only the Hermitian part of the Dirac mass term is independent which substantially simplifies the  task of testing experimentally the origin of neutrino mass. 
    We illustrate this program through some physical examples that allow simple analytical expressions.
   Our work shows that the Minimal Left-Right Symmetric Model is a self-contained theory of neutrino mass which can be in principle tested  at the LHC or the next hadron collider.
    
    \end{abstract}

\maketitle

%
%
\section{Introduction} 

 Understanding the origin of neutrino mass is a central task of the physics Beyond the Standard Model. After all, a non-vanishing neutrino mass is the only true failure of the SM and thus provides a fundamental window into the new physics. Over the years, the seesaw mechanism emerged as the main scenario behind the smallness of neutrino mass  \cite{Minkowski,Mohapatra:1979ia,rest}. By adding new neutral lepton singlets $N$ (one per generation) to the SM, and allowing for their gauge invariant masses $M_N$, one gets for the light neutrino mass matrix
\begin{equation}\label{seesaw-I}
M_{\nu}=-M_D^T\frac{1}{M_N}M_D
 \end{equation} 
where $M_D$ is the Dirac mass term between $\nu$ and $N$. The above formula  \eqref{seesaw-I} is as valid as long as $M_N \gg M_D$, a natural assumption for the gauge singlets $N$. 
In order to probe the seesaw mechanism, ideally one would have to find $M_D$ as a function of neutrino masses and mixing, namely of $M_\nu$, currently being probed in low energy experiments, and $M_N$, which could hopefully be determined at the LHC or future hadronic collider. Hereafter, this is what we will mean by disentangling the seesaw, and though in general it is not always guaranteed and should not be taken for granted a priori, an effective  disentangling would be analogous to the situation of charged fermions in which one can determine  their Yukawa couplings from the knowledge of their masses.

However, the situation arising from \eqref{seesaw-I} is given by (for an equivalent parametrization see  \cite{Casas:2001sr})
\begin{equation}\label{MD-I}
                                       M_D = i \sqrt{M_N} O \sqrt{M_{\nu}}                                                   
\end{equation}
where $O$ is an arbitrary complex orthogonal matrix
\begin{equation}
                                          O O^T = 1                                                                              
\end{equation}
 The arbitrariness of $M_D$ in \eqref{MD-I}  provides a blow to the program of probing the seesaw origin of the neutrino mass. Ironically,  the same arbitrariness is often used to make $M_D$ artificially large so that $N$ could be produced at the colliders (when $M_D$ vanishes $N$ are decoupled) but that is against the spirit of the seesaw as a way of obtaining small neutrino mass naturally.

A clarification is called for at this point. When we speak here of the origin of neutrino mass, we mean the physical  Higgs mechanism origin, as in the case of charged fermions and gauge bosons.  We do not have in mind the valid question of the values of these parameters. Before asking why the masses are what they are, we should first know where they come from. For example, a program aiming to solve for the mass hierarchies prior to the Higgs mechanism would have been obviously doomed from its very beginning. In that same spirit, we wish to have a theory that tells us in a verifiable and predictive manner whether neutrinos owe their masses to the Higgs mechanism and in which manner. Only then one can finally address the issue of the hierarchies of masses and mixings that we are all after.

In the case of charged fermions the Higgs origin of their masses implies the knowledge of Yukawa couplings from the values of masses $y_f \propto m_f/M_W$, and in turn allows to predict the Higgs decay rates into fermion - anti fermion pairs
\begin{equation}
   \Gamma (h \to \bar f f ) \propto m_h (m_f/M_W)^2                                                  
\end{equation}
Understanding the origin of neutrino mass is comparable to finding a theory that does  for neutrinos what the SM does for charged fermions, and in this sense the seesaw scenario by itself comes short.

This should not come out as a surprise. After all, the seesaw mechanism is an ad-hoc extension of the SM
and makes no attempt for a dynamical explanation of the V-A theory of weak interactions. This is to be contrasted with its left-right (LR) symmetric extension \cite{PatiSalam,pbreaking} that attributes the left-handed nature of weak interactions to the spontaneous breakdown of parity.
 The smoking gun signature of left-right symmetry is the existence of RH neutrinos ${\nu}_R$, leading to non-vanishing neutrino mass. 

In the minimal version of the theory, coined Minimal Left-Right Symmetric Model (MLRSM) based on extra Higgs scalars be LH and RH triplets \cite{Minkowski,Mohapatra:1979ia,MohSenj81}, the seesaw mechanism follows naturally from the spontaneous symmetry breaking with
\begin{equation}
                                               M_N\propto M_{W_R}                                                     
\end{equation}                                               
 where $M_{W_R}$ is the mass of the right-handed charged gauge boson. This offers a profound connection between the smallness of neutrino mass and the near maximality of parity violation in weak interactions \cite{Mohapatra:1979ia}. 

     In the MLRSM the left-right symmetry is broken spontaneously and can be either generalized parity $\mathcal{P}$ or generalized charge conjugation $\mathcal{C}$. 
      The impact of this symmetry on the properties of quarks and leptons is of fundamental importance. The case of $\mathcal{C}$ is easier to deal with since it leads to symmetric Dirac mass matrices of quarks and leptons, even after the spontaneous symmetry breaking. It implies  same LH and RH mixing angles in the quark sector, while, on the lepton sector the condition $M_D^T=M_D$ leads to the determination of $M_D$ which allows to disentangle the seesaw \cite{Nemevsek:2012iq}. This in turn allows, for example,  to predict the decay rates of $N \to W^+ e_L$ and $N \to h \nu$, and thus probe the Higgs origin of neutrino mass.

 The case of  $\mathcal{P}$ is however highly non-trivial since the originally Hermitian Dirac mass matrices lose this essential property after the symmetry breaking due to the emergence of complex vacuum expectation values. Instead, we have recently suggested an alternative approach of utilizing the decays of doubly charged scalars and  heavy SM doublet Higgs to probe $M_D$ \cite{Senjanovic:2016vxw}. We have also shown how to determine $M_D$ in the Hermitian case (unbroken parity in the Dirac Yukawa sector), which in a simple case of  the so-called type I seesaw and  same left and right leptonic mixing matrices takes the following unique form \cite{Senjanovic:2016vxw}
  \begin{equation}\label {MDirac}
M_D=i\, V_L \sqrt{m_{\nu}m_N}V_L^{\dagger}
\end{equation}
 This expression manifestly demonstrates the predictivity of the theory - all ambiguities are gone from the Dirac neutrino mass matrix.
 The Hermitian case may not be of  pure academic interest only; for light $W_R$  it may be a must due to a constraint of strong CP violation as we discuss in the following section, below \eqref{epsilonlimit}.

In this sequel of our paper \cite{Senjanovic:2016vxw}, we describe at length how $M_D$ can be computed in the case of being Hermitian, and we provide some simple appealing examples that can lead to transparent analytic expressions. We also elaborate on the phenomenological aspects of new particle decays and on the subsequent determination of $M_D$. The bottom line of our work is that, independently of whether $\mathcal{P}$ is broken or not in the Yukawa sector, the MLRSM is a self-contained theory of neutrino mass that allows for a direct probe of its Higgs origin. 

We should stress an important, essential aspect of our work. We make no assumption whatsoever regarding the breaking scale of the MLRSM, or equivalently the masses of $N$'s. Needless to say,  were they to be accessible at the LHC or the next hadron collider, this would allow us in principle to verify this program, but our work is more general than this scenario since it applies to any scale or any other imaginable way of knowing the masses and mixings of $N$'s.

The rest of the paper is organized as follows. In the following section we give the main features of the MLRSM, those which play an essential role in arriving at our main results, and set our formalism and notations. Its main purpose is to ease the reader's pain in following the technical aspects in later sections.  In section~\ref{section:lepton masses} we discuss the lepton masses, both charged and neutral, with a focus on the seesaw. This section plays a central role in understanding neutrino mass in the MLRSM and it could be useful even to the experts in the field. It has two subsections, the first being devoted to the parity conserving Yukawa sector that leads to Hermitian Dirac mass matrices. In this case we managed to solve analytically for $M_D$ as a function of $M_N$ and $M_\nu$. We give here a detailed expose' of this important result already found in~\cite{Senjanovic:2016vxw}. The general case is treated next in the second subsection, where we show that only the Hermitian part of $M_D$ is independent, and prepare the setup for phenomenological implications. The phenomenological analysis is left 
for  section \ref{section:pheno}, where we use a number of new processes, in particular the LNV violation decays of doubly charged scalars, to determine $M_D$ and thus demonstrate manifestly that the MLRSM is a self-contained theory of neutrino mass. Our conclusions and the outlook for future research are offered finally in section \ref{section:outlook}.

\section{Minimal Left-Right Symmetric Model}\label{section:mlrsm}

The MLRSM is based on the following symmetry group 
\begin{equation}\label{LRgroup}
              \mathcal{G}_{LR} = SU(2)_L \times SU(2)_R \times U(1)_{B-L} 
              \end{equation}
where on top of the LR symmetric gauge group, a discrete generalized parity  $\mathcal{P}$ ensures a LR symmetric world prior to spontaneous symmetry breaking. 

We focus on the leptonic sector only - for the quark sector see \cite{Senjanovic:2014pva,Senjanovic:2015yea}. Under \eqref{LRgroup} the leptonic fields transform as
\begin{equation}
  \ell_{L,R} = \left( \begin{array}{c} \nu \\ e \end{array}\right)_{L,R}.                              
         \end{equation}
and under $\mathcal{P}$ as
\begin{equation}
             \ell_L \leftrightarrow \ell_R.                                    
\end{equation}

The new Higgs sector consists of left and right $SU(2)$ triplets
  $ \Delta_L (3,1,2)$ and $\Delta_R (1,3,2)$,   respectively,                     
where the quantum numbers denote the representation content under \eqref{LRgroup}. The RH triplet $\Delta_R$ is responsible for the breaking of $G_{LR}$ down to the SM gauge symmetry, and its non-vanishing vev $v_R$ (notice that it can be made real) gives masses to the new heavy gauge bosons $W_R$ and $Z_R$ and the RH neutrinos $N$.

The field decomposition of the triplets has the following form     
\begin{equation}
\Delta_{L,R}=\left(
 \begin{array}{c c}
  \delta_{L,R}^+ /\sqrt{2}& \delta_{L,R}^{++} \\ [3pt]
\delta_{L,R}^0 & - \delta_{L,R}^+ /\sqrt{2}
\end{array} 
\right)
\end{equation}
Besides the usual singly charged fields  $\delta_L^{\pm}$ ($\delta_R^{\pm}$ gets eaten by the $W_R^{\pm}$ fields), the doubly charged states $\delta_{L,R}^{++}$ play an important role in lepton number violating decays and in determining $M_D$. As in the original work \cite{pbreaking}, at the first stage of symmetry breaking $v_L=\langle \delta_L^0\rangle= 0, v_R=\langle \delta_R^0\rangle\neq 0$.

On top of the new Higgs multiplets, there must also exist a $SU(2)_L \times SU(2)_R$ bi-doublet, containing the usual SM Higgs field, $\Phi (2,2,0)$ with the decomposition
\begin{equation}
\Phi= \left[\phi_1, i\sigma_2 \phi_2^*\right],\quad \phi_i= \left( \begin{array}{c} \phi_i^0 \\ \phi_i^- \end{array}\right),\quad i=1,2. 
\end{equation}
 The most general vev of 
$\Phi$ can be written as
\begin{equation}
\langle\Phi\rangle=v\, \text{diag} (\cos\beta,-\sin\beta e^{-ia})
\end{equation}
 Under parity one has as 
\begin{equation}\label{parity}
 \Delta_L\leftrightarrow \Delta_R,\quad \Phi\rightarrow\Phi^{\dagger}
\end{equation}
The leptonic Yukawa interaction is given by
\begin{equation}\label{Ly}
\begin{split}
\mathcal{L}_Y= & \,-\overline{\ell_L}(Y_1\Phi-Y_2 \sigma_2 {\Phi}^* \sigma_2) \ell_R \,\\
&-\frac{1}{2}\left(\ell_L^TY_L i\sigma_2\Delta_L\ell_L+ \ell_R^TY_R i\sigma_2\Delta_R\ell_R\right)+\text{h.c.}
\end{split}
\end{equation}
so that because of  \eqref{parity} one has
\begin{equation}\label{Yparity}
Y_{1,2}=Y_{1,2}^{\dagger}, \quad Y_L=Y_R
\end{equation}
These relations are central to our discussion and to our main results. 

Introduce next $N_L= C \bar \nu_R^{T}$, which from \eqref{Ly} leads immediately to the heavy neutrino mass matrix 
\begin{equation}
M_N=v_R Y_R^*
\end{equation}

The SM Higgs doublet $h$ and the new heavy doublet $H$ are the linear combination of $\phi_i$
\begin{equation}
h = c_\beta \phi_1 + e^{-ia} s_\beta \phi_2,  \quad   H = - e^{ia} s_\beta \phi_1 + c_\beta \phi_2,
\end{equation}
where $c_\beta \equiv \cos_\beta, s_\beta \equiv \sin_\beta$ hereafter.  The new doublet $H$ is basically decoupled, i.e., out of the LHC reach, since it leads directly at the tree level to the flavor violation in the K and B-meson sectors. This implies a lower limit $m_H   \gtrsim 20 \text{ TeV}$ ~\cite{OtherLR, Bertolini:2014sua}, which is far above the LHC reach (for a recent study regarding the future hadron collider, see~\cite{Dev:2016dja}).

It is useful to rewrite the Yukawa interaction of the bi-doublet $\Phi$ as the function of the physical fields $h$ and $H$ and the charged lepton and neutrino Dirac mass matrices 
\begin{equation}\label{Phi-int}
\begin{split}
\mathcal{L}_\Phi = & \,- \overline{\ell}_L \left[\frac {M_D^\dagger}{v} h  - \frac{M_e +e^{-ia} s_{2\beta} M_D^\dagger}{v c_{2\beta}} H\right] N_R\\
&\, + \overline{\ell}_L \left[\frac {M_e}{v} i\sigma_2 h^*  - \frac{M_D^\dagger +e^{ia} s_{2\beta} M_e }{v c_{2\beta}} i\sigma_2  H^*\right] e_R
\end{split}
\end{equation}
where $M_e$ and $M_D$ are the charged lepton and neutrino Dirac mass matrices, respectively, and are  given by
  \begin{align} \label{Md-eq}
M_D&=-v ( c_{\beta} Y_1+ e^{-i a}s_{\beta} Y_2)\\ 
\label{Me-eq} 
M_e &=  v (e^{-i a} s_{\beta} Y_1+ c_{\beta} Y_2).
\end{align}
The measure of spontaneous CP violation is provided by the small parameter $s_a t_{2\beta}$ which can be shown to satisfy \cite{Senjanovic:2014pva}
\begin{equation}\label{epsilonlimit}
s_a t_{2\beta}\lesssim \frac{2m_b}{m_t}
\end{equation}
It can be shown that the same parameter measures the difference between  right and left-handed quark mixing matrix and  thus controls the weak contribution to the strong CP violating parameter $\bar \theta$. For light $W_R$ one has $s_a t_{2\beta}$ is practically vanishing \cite{Maiezza:2014ala} in order to keep $\bar \theta$ acceptably small. The point is that with the spontaneously broken parity the strong CP parameter $\bar \theta$ is finite and calculable in perturbation theory~\cite{Beg:1978mt}.

Once $\langle \Phi \rangle$ is turned on, the left-handed triplet $\Delta_L$ gets a small induced vev
$v_L\propto v^2 /v_R$
providing a hierarchy of $SU(2)_L$ breaking \cite{MohSenj81}. It is important to stress that $v_L$ is naturally small, since it is protected by a symmetry \cite{MohSenj81} (for a recent discussion, see \cite{Maiezza:2016ybz}). The small $v_L$ is thus a direct source of neutrino mass, the so-called type II seesaw. It is interesting to contrast this situation with the usual type II SM scenario where one adds ad-hoc a $SU(2)$ triplet in order to give neutrino a non-vanishing mass  \cite{typeII}. The latter case is an example of an a posteriori model building, while in the MLRSM this is a result of the underlying structure of the theory. Just as the type I seesaw emerges naturally in this theory, since RH neutrinos are a must, the same mechanism that gives them Majorana masses
leads automatically to the direct type II contribution to light neutrino masses.

\section{Lepton Masses}
\label{section:lepton masses}

This is the central section of our work. We go here from the weak to the mass basis, which requires some care due to the common source of charged lepton and neutrino masses in the MLRSM.

   The charged lepton mass matrix can be diagonalized by performing unitary transformations $E_{L}$ and $E_R$ on the LH and RH charged lepton fields, respectively
\begin{equation}
M_e = E_L m_e E_R^\dagger
\end{equation} 
More precisely, one rotates the LH and RH doublets 
\begin{equation}\label{originalrot}
\ell_L  \rightarrow E_L \ell_L,  \,\,\,\,\,\,\,\, \ell_R \rightarrow E_R \ell_R
\end{equation} 
so that at this point the gauge interactions of charged gauge bosons remain still diagonal. The leptonic mixings, i.e. the PMNS matrix $V_L$ and its right-handed analog $V_R$ will then be simply the unitary transformations that diagonalize LH and RH neutrino mass matrices, respectively. 

Since parity is broken by the complex vev  $\langle\Phi\rangle$, in general $E_L \neq E_R$, and thus  
\begin{equation}
U_e = E_R^{\dagger} E_L
\end{equation} 
provides a measure of parity breaking. 

Equation \eqref{originalrot} implies a redefinition of the Dirac neutrino mass matrix
\begin{equation}
 M_D \rightarrow E_R M_D E_L^\dagger 
\end{equation}
From \eqref{Md-eq}  and  \eqref{Me-eq} it follows that
\begin{align}
\label{eq1}
 &M_D - U_{e}M_D^{\dagger}U_{e}=is_at_{2\beta}(e^{ia}t_{\beta}M_D+m_{e})
 \\[8pt]
 \label{eq2}
  &m_{e}-U_{e} m_{e}U_{e}=-is_at_{2\beta}(M_D+e^{-ia}t_{\beta}m_{e})
 \end{align}
  From the above equations it is clear that $U_e$ is actually a function of $M_D$ and vice versa as discussed below. 

From the Yukawa interaction and \eqref{Yparity}, one gets for the $(\nu_{L}, N_{L})$ mass matrix 
\begin{equation}\label{nuNmassmatrix}
\left( \begin{array}{c c}\dfrac{v_L}{v_R} U_e^TM_N^*U_e & M_D^T \\[10pt] M_D & M_N \end{array}\right)  
\end{equation}
This is a combination of both type II and type I seesaw matrices, with the important proviso of $U_e$
entering in the direct type II mass term. In other words, in the MLRSM not only $M_D$ enters the neutrino mass, but also $U_e$ - this point is typically missed in the literature.

Under the usual seesaw assumption $M_N\gg M_D$ the matrix (20) can  be readily block-diagonalized, through the $\nu-N$ mixing,  to the leading order in $M_D/M_N$ given by 
\begin{equation}
 \left( \begin{array}{c} \nu \\ N \end{array}\right)_L \rightarrow 
 \left( \begin{array}{c c}  1 &  \Theta^{\dagger}   \\  -\Theta & 1   \end{array}\right)   \left( \begin{array}{c} \nu \\ N \end{array}\right)_L
\end{equation}
with
\begin{equation}
\Theta=\frac{1}{M_N}M_D
\end{equation}
This in turn leads to the celebrated seesaw expression for the neutrino mass
  \begin{equation}\label{seesaw}
M_{\nu}=\frac{v_L}{v_R}U_{e}^TM_N^*U_{e} -M_D^T\frac{1}{M_N}M_D
\end{equation}
  Thus, at the leading level $M_\nu$ and $M_N$ stand for the approximate (Majorana) mass matrices for the light and heavy neutrinos, respectively. The neutrino mass matrix is given as a function of $M_D$ and $M_N$. 
   Together with \eqref{eq1} and \eqref{eq2}, the above equation serves to compute $M_D$ and thus to disentangle the seesaw.

  The next step is then to diagonalize these matrices by the unitary rotations $V_L$ and $V_R$ respectively. From \eqref{originalrot} one gets immediately
\begin{equation}
M_{\nu}=V_L^*m_{\nu}V_L^{\dagger}
\end{equation}
where $m_\nu$ stands for the diagonal matrix of the light neutrino masses and $V_L$ is the standard PMNS mixing matrix, which amounts to
  \begin{equation}\label{Lrot}
\nu_{L} \rightarrow V_L\nu_L
\end{equation}

Similarly, $M_N$ is diagonalized by the unitary $V_R$
       \begin{equation}
M_N=V_Rm_NV_R^T
\end{equation}
where $m_N$ stands for the diagonal matrix of the heavy neutrino masses. This means equivalently
\begin{equation}\label{Rrot}
N_{L} \rightarrow V_R^* N_L
\end{equation}
when going from the weak to the mass basis. The apparent difference in the rotations among the light and heavy neutrino masses is due to the fact that $N$ corresponds to complex conjugate fields of the right-handed neutrinos $\nu_R$, so that one has $\nu_{R} \rightarrow V_R \nu_R$ in complete analogy with \eqref{Lrot} .

It is easy to see that \eqref{Lrot} and \eqref{Rrot} give the usual form of the charged weak interaction
    \begin{equation}
 \mathcal L_{W}= - \frac{g}{\sqrt 2} \left( 
  \overline \nu_L V_{L}^\dagger  \slashed{W}_{\!L} e_L + 
  \overline N_R  V_{R}^\dagger  \slashed{W}_{\!R} e_R\right) + \text{h.c.}
\end{equation}
so that $V_R$ is the right-handed counterpart of the PMNS matrix, making clear the choices of the $V_L$ and $V_R$ rotations.

  Before we enter into the nitty-gritty of our program, we give a simple example where actually the canonical type I seesaw becomes negligible - and yet, interestingly enough, both $M_D$ and $U_e$ are calculable, as it turns out, by the charged lepton masses and the mixing matrices $V_L$ and $V_R$.  In this case, the term $v_L$ by definition dominates the contribution to neutrino masses and from the \eqref{seesaw} one gets
\begin{equation}
m_{\nu}=\frac{|v_L|}{v_R} m_N, \quad U_e= e^{-i \theta_L}V_RV_L^{\dagger}
\end{equation}
where we have used $v_L=|v_L|e^{-i \theta_L}$.  Thus, the Dirac masses can be determined from \eqref{eq2} as a function of the  difference of LH and RH leptonic mixing matrices and the charged lepton masses
\begin{equation}
M_D=\frac{  e^{-2i \theta_L}V_RV_L^{\dagger} m_e V_RV_L^{\dagger}-m_e}{i s_at_{2\beta}} -e^{-i a}t_{\beta}m_e
\end{equation}
It is noteworthy that the Hermitian limit $s_a t_{2\beta}\simeq0$ is not smooth and it has to be dealt with carefully, as we discuss in the next section. This case exemplifies the fact that the theory predicts $M_D$, and it does it even when the contribution of $M_D$ to neutrino masses is small. 

\subsection{The tale of unbroken parity}\label{unbrokenP}

What would happen if parity was not broken by the  vev of $ \Phi$, i.e., what if the small parameter $s_a t_{2\beta}$ was to be negligible? Could $M_D$ be found analytically? The answer is yes as we show now.

From \eqref{eq1} one would have 
\begin{equation}
M_D=M_D^{\dagger}, \quad U_e=  \mathbb{I}  \,\,(\text{up to signs)}
\end{equation}
By taking the complex conjugate of  \eqref{seesaw} and dividing on the left and right by $\sqrt{M_N}$, one readily obtains an equation between symmetric matrices
\begin{equation}\label{HHT}
H H^T=\frac{v_{L}^*}{v_R}-\frac{1}{\sqrt{M_N}}M_{\nu}^*\frac{1}{\sqrt{M_N}}
\end{equation}
where $H$ is a Hermitian matrix defined as
 \begin{equation}\label{MD-H}
H=\frac{1}{\sqrt{M_N}}M_D\frac{1}{\sqrt{M_N^*}}
\end{equation}
Since $\text{Im}\,\text{Tr}\,\left(HH^T\right)^n = 0$ for any $n$ and Hermitian $H$, 
one has the following conditions
 \begin{equation}\label{conditions}
  \text{Im}\,\text{Tr}\left[ \frac{v_{L}^*}{v_R}-\frac{1}{M_N}  M_{\nu}^*\right]^n=0, \quad n=1,2,3.
\end{equation}

It turns out
that the equations (26) and (27) allow for a direct determination of $M_D$.
The crucial  step is to decompose the symmetric matrix \eqref{HHT} as
  \begin{align}\label{Omatrix}
H H^T= O s O^T
  \end{align}
where $O$ is a complex orthogonal matrix and $s$ is known as the symmetric normal form \cite{Gantmacher}, to be determined from the knowledge of neutrino masses and mixing. 
This is an unorthodox method for particle physicists, since one normally uses a unitary matrix instead of  orthogonal, since it guarantees the diagonalization of a symmetric matrix we are dealing with. However, the unitary matrix approach is not suitable for our task.
  In this case, however, the form $s$ is not guaranteed to be diagonal.

From \eqref{Omatrix}, when $H$ is real and symmetric, it follows immediately that $H$ becomes $O \sqrt s O^T$. It can be shown that in the general complex case this expression generalizes to 
\begin{equation}\label{detH}
H=O\sqrt{s}EO^{\dagger} 
\end{equation}
The condition $H=H^{\dagger}$ now becomes 
 \begin{equation}\label{Hfound}
\sqrt{s}E=E\sqrt{s^{*}}, \quad \, E^T=E^*=E^{-1}
\end{equation}

Using \eqref{MD-H} and \eqref{detH},  one can achieve the task of disentangling the seesaw by determining  $M_D$   as     
\begin{equation}
M_D=\sqrt{M_N}\,O\sqrt{s}EO^{\dagger}\sqrt{M_N^*}
\end{equation}
Since $O$, $s$ and $E$ all follow from the knowledge of $M_\nu$ and $M_N$,  this manifestly shows how in the parity conserving case $M_D$ can be determined from the knowledge of light and heavy neutrino masses and mixings. 

The above expression is valid for any $M_\nu$ and $M_N$, i.e., any normal form $s$. It is illustrative to focus on the situation when $s$ takes a diagonal form, in which case the constraints \eqref{conditions} imply only two distinct possibilities
\begin{equation}\label{s-diagonal}
  s_{I}=\text{diag}(s_1,s_0,s_2) ,\quad   s_{II}=\text{diag}(s,s_0,s^*) 
    \end{equation}
with $s_{0,1,2}$ belonging to $\mathbb{R}$. The matrix $E$ is  found to be
  \begin{equation}
E_{\text{I}}=  \left(   \begin{array}{ccc}
 1& 0&0 \\ 
  0& 1 & 0 \\ 
0 & 0 & 1
\end{array}   \right), \qquad  E_{II}=  \left(   \begin{array}{ccc}
 0& 0&1 \\ 
  0& 1 & 0 \\ 
1 & 0 &  0
\end{array}   \right)
    \end{equation}
for the respective values $s_I$ and $s_{II}$.
Equation \eqref{s-diagonal} can be generalized to any number of generations $n$: for $n$ even, for every eigenvalue $z$, there is also an eigenvalue $z^*$.
For $n$ odd there is on top one real eigenvalue. The matrix $E$ in this case has a 1 in the diagonal  for each corresponding 
real eigenvalues and two 1's  symmetrically opposed in the anti-diagonal for each corresponding complex eigenvalue and its conjugate. 

A comment is called for. In  \cite{Senjanovic:2016vxw} we have chosen a different decomposition of $M_D$
\begin{equation}
M_D= V_R \sqrt{m_N}H' \,\sqrt{m_N} V_R^{\dagger}  
\end{equation}
 which led to  a different decomposition of  $H'H'^T$ in terms of $O'$ and $s'$ in full analogy with \eqref{Omatrix}.
It is straightforward to show that 
\begin{equation}
O'=\frac{1}{\sqrt{m_N}}V_R^{\dagger}\sqrt{M_N}\,O, \quad s'=s
\end{equation}

Since in general $M_N$ and $M_\nu$ are arbitrary complex matrices up to constraints \eqref{conditions}, no general analytic expression can be offered for $M_D$.  The following examples may help illustrate what is going on.

(i) $V_R = V_L$. Imagine an idealized situation with parity unbroken in the leptonic sector. Clearly, $O' = 1$ and thus
 \begin{equation}
O= \frac{1}{\sqrt{M_N}}V_L\sqrt{m_N},\quad s=\frac{v_L}{v_R}-\frac{m_{\nu}}{m_N}
\end{equation}
and 
\begin{equation}\label{MD-VL=VR}
M_D= V_L m_N \sqrt{\frac{v_L}{v_R}-\frac{m_{\nu}}{m_N}}V_L^{\dagger}
\end{equation}
which for $v_L=0$, i.e., in the type I seesaw limit, gives the simple form of \eqref{MDirac}. Matrix \eqref{MD-VL=VR} is defined up to signs since, strictly speaking, $U_e$ is defined up to signs.
This example is to be contrasted with the situation in the SM seesaw scenario, in which $M_D$ is plagued by the arbitrariness of a complex orthogonal matrix, here fixed completely.

(ii) Case of two generations. It is always illustrative to imagine a two-generation world where one can offer simple analytic formulas. We focus on a physically appealing situation of non-degenerate neutrinos, in which case the form $s$ becomes diagonal. For simplicity, choose again $v_L = 0$ so that the constraints \eqref{conditions} become
 \begin{equation}
  \text{Im}\,\text{Tr}\,M_N^{-1} M_{\nu}^* =0, \quad  \text{Im}\,\text{det}\, M_N^{-1} M_{\nu}^* =0
  \end{equation}
We can readily write down the eigenvalues of $s=\text{diag}(s_{-},s_{+})$ 
\begin{equation}
s_{\pm}= \tfrac{1}{2} \text{Tr} \,M_N^{-1} M_{\nu}^* \ \pm  \sqrt{\tfrac{1}{4}\!\left(\text{Tr}\, M_N^{-1} M_{\nu}^*\right)^2 -\text{det}  M_N^{-1} M_{\nu}^* }
\end{equation}
which shows explicitly that $s_{\pm}$ are either both real or form a complex conjugate pair. In turn, by writing
\begin{equation}
O=\left(
\begin{array}{rl}
 \cos \theta & \sin \theta\\
  -\sin \theta & \cos \theta \end{array}
  \right)
\end{equation}
one gets
\begin{equation}
\sin 2\theta= \frac{ \Big(\sqrt{M_N^{-1}}M_{\nu}^* \sqrt{M_N^{-1}}\Big)_{12} }{  \sqrt{\tfrac{1}{4}\!\left(\text{Tr}\, M_N^{-1} M_{\nu}^*\right)^2 -\text{det}  M_N^{-1} M_{\nu}^* }  }
\end{equation}

We now turn to the general case, with one last comment regarding the Hermitian $M_D$. In this case, the freedom in the RH quark mixing is all gone, and the constraints from the $K$ and $B$ meson system imply $W_R$ too heavy to be accessible at the LHC~\cite{Bertolini:2014sua}. The parity conserving situation is then automatically postponed to a future hadron collider.

 \subsection{Broken parity: setting the stage} \label{setting stage}

We have seen in the previous section how a parity conserving situation with $M_D$ Hermitian allows for its  determination as a function of $M_N$ and $M_\nu$. This is in complete analogy with the SM situation where a knowledge of charged fermion masses fixes their Yukawa couplings and allows to predict branching ratios for the Higgs boson decays. In the present case one needs to know both $M_{\nu}$ and $M_N$ due to their Majorana nature, but still, $M_D$ is then uniquely fixed and we can determine  associated decay rates such as $N \to h \nu$, $N \to Z \nu$ and $N \to W \ell$.

 When parity gets broken, however, we have not (yet) managed to compute 
 $M_D$  Not all is lost though, as we have argued in~\cite{Senjanovic:2016vxw}. The crucial point is the existence of a number of physical processes, in particular the same sign leptonic decays of doubly charged scalars $\delta_{L,R}^{\pm}$, that depend crucially on $M_D$ and can serve to verify the Higgs seesaw origin of neutrino mass. We discuss these processes in the next section. 

Now, although we found no way of computing $M_D$ analytically in the general case, we show that only its Hermitian part is independent, which makes the predictions significantly easier to test. As a first step, we compute the matrix $U_e$ as a function of $M_D$.
Multiplying \eqref{eq2} with $m_e$ and taking the square root gives
\begin{align}\label{Ulformula}
U_{e}&=\frac{1}{m_{e}}\sqrt{m_{e}^2+is_at_{2\beta} (t_{\beta}e^{-ia}m_{e}^2+m_{e}M_D)} 
\end{align}
 We keep a small term $s_a t_{2\beta} m_e^2$ in order to emphasise that \eqref{Ulformula} is exact.  Notice the fact that $U_e U_e^\dagger =1$ reduces by half the number of independent elements of $M_D$, implying  that the anti-Hermitian part $M_D^A = \frac{1}{2}(M_D - M_D^\dagger)$ becomes a  function of the Hermitian part $M_D^H = \frac{1}{2}(M_D + M_D^\dagger)$. This can be shown from \eqref{eq1}. 
 
 Here we give the leading expression in $s_a t_{2\beta}$, based on the following  expansion of the square root of a matrix (see Appendix A in \cite{Senjanovic:2015yea})
    \begin{equation}
\left(\sqrt{m^2+ i \epsilon A^2}\right)_{ij}=m_i\delta_{ij}+i \epsilon \frac{A_{ij}}{m_i+m_j}+O(\epsilon^2)
\end{equation}
Using this in \eqref{Ulformula} gives
\begin{equation} \label{U-e1}
(U_{e})_{ij}=\delta_{ij}+i s_a t_{2\beta}\left[ \frac{t_{\beta}\delta_{ij}}{2} +(\mathcal{H_D})_{ij} \right]+O(s_a^2 t_{2\beta}^2)
\end{equation}
where we have defined
\begin{equation}\label{HD-matrix}
(\mathcal{H}_D)_{ij}= \frac{\,\,\,(M_D^H)_{ij}}{m_{e_i}+m_{e_j}}.
\end{equation}

From \eqref{eq1} and \eqref{U-e1} it follows readily
\begin{align}\label{antihermitianMD}
M_D^{A}\!=\!\frac{is_at_{2\beta}}{2}\!\Big(\! m_{e}\!+\!2t_{\beta}M_D^H\!+\!\mathcal{H_D}M_D^H\!+\!M_D^H\mathcal{H_D} \!\Big) \!+\! O(s_a^3t_{2\beta}^3)
\end{align}
This is an important expression that says that only $M_D^H$ is physical, halving the effective degrees of freedom in the task of probing the seesaw. The elements of $M_D^H$ can be determined numerically from the seesaw, and we address this in a forthcoming publication.

\subsection{Broken parity: general situation}

As we argued, the expression for $U_e$ in \eqref{Ulformula} shows manifestly its dependence on $M_D$. The trouble is that this $U_e$ is not automatically Hermitian and thus becomes really useful only when expanded in small $s_at_{2\beta}$. What one needs is an explicit unitary expression for $U_e$, valid to all orders in $s_at_{2\beta}$ which we now provide. 

 It turns out useful to rewrite equations \eqref{eq1} and \eqref{eq2} as
  \begin{align}
\label{eq1x}
 & U_{e}M^{\dagger} U_{e}-M=0
 \\
 \label{eq2x}
  &U_{e} m_{e}U_{e} - m_{e}=i s_a t_{2\beta} M
 \end{align}
where $M$ is given by
  \begin{align}\label{Mgiven}
M =   M_D +e^{-ia}t_{\beta}m_{e}
\end{align}
We leave as an exercise for the reader to show from the above equations can be written as
 \begin{align}\label{Ueunitary} 
& M=\sqrt{MM^{\dagger}} U_e \\ 
  & U_{e}\!=\!\frac{1}{\sqrt{\!\Big(\!m_e\!+\frac{i s_a t_{2\beta}}{2} M\! \Big)\! \Big(\!m_e\!-\frac{i s_a t_{2\beta}}{2} M^{\dagger}\!\Big)} }\Big(\!m_e\!+\!\tfrac{i s_a t_{2\beta}}{2} M\!\Big)\label{holygrail}
 \end{align}
The last equation is what we were after and what we promised: an explicitly unitary form of $U_e$ as a function of $M_D$.  It becomes explicitly unity when $s_a t_{2\beta}$ vanishes. 

There is actually more in these equations; they tell us finally what is really going on. From \eqref{Ueunitary} the matrix $U_e$ can be viewed as the ``phase'' of $M$, and \eqref{holygrail} shows that due to parity this phase is determined in terms of $M$ itself. 
In other words,  in spite of being broken spontaneously, $\mathcal{P}$ still acts and sets the phase of $M$ equal to the phase of  $m_e+ \frac {i}{2} s_a t_{2\beta} M$. 

This is seen nicely in the one-generation toy example when
the matrix $M$  is just a complex number. From \eqref{Mgiven} one then has $m_D=\rho e^{i \theta}-e^{-ia}t_{\beta}m_e$ and \eqref{holygrail} shows that $\theta$ is not arbitrary, but actually a function of $\rho$ itself: $ \sin {\theta}=s_at_{2\beta} \,\rho/2 m_e$.
Parity does its job  by halving the number of independent degrees of freedom. Notice also that $m_D$ is bounded from above and below, as expected from \eqref{epsilonlimit} and an analogy with the quark system.

 A word on the parametrization of $M_D$. Instead of parametrizing the Hermitian degrees of freedom of $M_D$  by its Hermitian part as we did above in \ref{setting stage},  one could have used the Hermitian matrix  $\mathcal{M}=\sqrt{M M^{\dagger}}$ as well, without any loss of generality. One has then from \eqref{Ueunitary} and  \eqref{holygrail}
\begin{align}
&(U_{\ell})_{ij}= \delta_{ij}+is_at_{2\beta}  \frac{\mathcal{M}_{ij}}{m_{e_i}+m_{e_j}} +O(s_a^2t_{2\beta}^2)\\
&(M_D+e^{-ia}t_{\beta}m_e)_{ij} =\mathcal{M}_{ij}+is_at_{2\beta}  \frac{\mathcal{M}_{ik}\mathcal{M}_{kj}}{m_{e_k}+m_{e_j}}+O(s_a^2t_{2\beta}^2)
\end{align} 
The task now becomes to determine $\mathcal{M}$ from the seesaw. Again, we have found no analytic form yet, but one can always use a numerical procedure. It can however be illustrated on our toy one-generation example, where one has readily for $m_D=|m_D|e^{i\theta_D}$ 
\begin{align}
m_D= i \sqrt{m_{\nu}m_N} V_R 
\end{align}
The memory of parity provides an important constraint, written here to the leading order in 
$s_at_{2\beta}$ 
\begin{align}
 \theta_D=\frac{s_a t_{2\beta}}{2}  \left[ \pm\, 2 t_{\beta} +\frac{m_e}{\sqrt{m_{\nu}m_N}}+\frac{\sqrt{m_{\nu}m_N}}{m_e} \right]
\end{align} 

A comment is called for. The reader familiar with the RH quark mixing in the MLRSM can recognize here the expression given in the formula (A11) of~ \cite{Senjanovic:2015yea}, with the understand that one here we have used the seesaw formula for $|m_D|$. The reason is simple: were neutrinos Dirac particles, one would have a complete analogy between the leptonic and quark sectors and $\theta_D$ would correspond to the conjugate of the RH quark phase. An interested reader can also find an exact expression for $\theta_D$ analogous to  (C1) of Appendix C in ~\cite{Senjanovic:2015yea}, true to all orders in $s_at_{2\beta}$.
 
 It is worthwhile to confront the MLRSM one-generation case with the SM seesaw scenario. There the phase of $m_D$ is simply not physical since its impact can be countered by the phase of the electron field. Here, on the contrary, the phase of $m_D$ is physical (see e.g. \eqref{Htoeebar} in the next section) and related to the phase of $V_R$ - and furthermore it can be computed as a function of electron and neutrino masses, and the vevs of the bi-doublet. 

The final message is clear. Whatever parametrization one chooses to use, the important point is that $\mathcal{P}$ fixes $n^2$ elements of $M_D$. We can say that the spontaneous breaking of parity plays an equally important role as the seesaw itself in determining $M_D$.

The crucial point is all this is the softness of the spontaneous symmetry breaking which makes the theory remember that parity was there to start with. One is tempted to agree with Coleman on calling it a hidden, rather than spontaneously broken symmetry.

\section{Phenomenological implications}
\label{section:pheno}

  We now turn our attention to the phenomenological issues, and we do it in full generality without assuming Hermitian $M_D$.  As stressed before, besides eliminating the freedom in $M_D$, the MLRSM offers a number of new physical processes which can pave the road for probing of $M_D$ and the origin of neutrino mass.

    First of all, there is an exciting possibility of observing direct Lepton Number Violation in the $W_R$ decays due to the Majorana nature of RH neutrinos $N$. Namely, once produced $W_R$ decays either into two jets or charged leptons and $N$'s which then further decay into charged leptons and two jets. The main semi-leptonic decays consist then equally of same and opposite sign charged lepton pairs accompanied by the pairs of jets. The former provide a direct LNV, and together with the latter allow for a unique direct test of the Majorana nature of $N$. The same sign process \cite{Keung:1983uu}, coined Keung-Senjanovic (KS),  is a high-energy hadron collider analog of the neutrinoless double beta decay, with a clear signature that the outgoing charged leptons have RH chiralities. It has been argued that the charged lepton chirality can actually be measured at the LHC~\cite{Ferrari:2000sp,Han:2013}. In the MLRSM there is also a deep connection between the neutrinoless double beta decay and the high-energy KS process, studied in~\cite{Tello:2010am}.

 \subsection{Decays and the probe of $M_D$} \label{decaysthroughMD}

  The KS signature, if accessible at the LHC or the next hadron collider, would allow for the determination of $M_N$ (for phenomenological studies, see e.g.~\cite{Das:2012ii}), i.e., $m_N$ and the leptonic RH mixing matrix $V_R$, the first step towards the probe of the seesaw origin of neutrino mass. Next, through $\nu - N$ mixing induced by the non-vanishing $M_D$ elements, $N$ can decay into LH charged leptons too with the following rate
\begin{equation}\label{N2Wdecay}
   \begin{split}
&\Gamma (N_i \to W_L^+ e^{}_{L_{ j}})\propto  \frac{  m_{N_i}}{M_{W_L}^2}
  \big|(V_R^{\dagger} M_D)_{ij}\big|^2
\end{split}
 \end{equation}
 In the above and hereafter, we will not worry about the precise rates; we give flavour and mass dependence up to overall dimensionless constants.
 
In the same manner,  decays such as $N \to Z \nu$ and $N \to h \nu$ (we are assuming $N$ to be also heavier than $Z$ and $h$, otherwise the opposite happens) are less exciting since they involve missing energy. We give anyway their decay rates for the sake of completeness
\begin{equation}
   \begin{split}
&\Gamma (N_i \to Z \nu_{ j}) \propto \Gamma (N_i \to h \nu_{ j}) \propto  \frac{  m_{N_i}}{M_{W_L}^2}
  \big|(V_R^{\dagger} M_D V_L)_{ij}\big|^2
\end{split}
 \end{equation}
 These processes, especially $N \to W^+ e_L$, can serve in the determination of $M_D$; for phenomenological studies see~\cite{Nemevsek:2012iq, Chen:2013fna}.

In the SM seesaw scenario these decays would of course also happen~\cite{Buchmuller:1991tu}, with an important difference being that  $M_D$ is ambiguous there. Worse, the production of $N$ in the SM seesaw can only be achieved through $M_D$, which requires $M_D$ to be large, contrary to the seesaw philosophy of $M_N \gg M_D$ as an explanation of the smallness of neutrino masses.
Moreover, in the MLRSM there are other fundamental processes which can probe $M_D$, discussed in what follows.
\\
\\
{  $\bullet$ Decays of the  doubly charged scalars $\delta_{L,R}^{++}$.}
 \vspace{0.1cm}
\\
 Doubly charged scalars are produced pairwise by the Z-boson and the photon, and unless are very light, are expected to be less accessible than the $W_R$. Nonetheless, their lepton number violating decays play an important role in disentangling the seasaw. From (12), the relevant Yukawa interaction is
     \begin{equation}
-\mathcal L_{\delta}\!=\!\frac{1}{2}   \, \delta_L^{^{++}}  e_{L}^T   \bigg(\!U_{e}^T \frac{M_N^*}{v_R} U_e\!\bigg)  e_L\!+ \frac{1}{2}\delta_R^{^{++}}  e_{R}^T \bigg(\!  \frac{ M_N^*}{v_R}\! \bigg)  e_R \!  
    \end{equation}
The matrix $U_e$ provides the expected mismatch between LH and RH states. Notice first that 
$\delta_R^{--} \rightarrow e_R e_R$ decays measure, complementary to the KS process, the masses and mixings of $N$'s
\begin{equation}
\Gamma({\delta_R^{--} \to e_{R_i}e_{R_j}}) \propto \frac{m_{\delta_R^{--}}}{M_{W_R}^2} \big|( M_N)_{ij} \big|^2
\end{equation}
These decays have been studied recently at the lepton colliders in~\cite{Dev:2018upe}.

 The LH analog decays play an even more important role in verifying the seesaw mechanism due to the presence of $U_e$ matrix, since one has 
\begin{equation}
\Gamma({\delta_L^{--} \to e_{L_i}e_{L_j}} )\propto \frac{m_{\delta_L^{--}}}{M_{W_R}^2} \big|\big(U_{e}^{\dagger} M_N U_e^*\big)_{ij} \big|^2
\end{equation}
Up to a proportionality factor, this is precisely the direct type II contribution to neutrino mass, i.e., 
the upper-left block of the neutrino mass matrix in \eqref{nuNmassmatrix}.  These decays can thus directly probe the pure type II seesaw as discussed in~\cite{Kadastik:2007yd} since in that case one has 
$\Gamma({\delta_L^{--} \to e_{L_i}e_{L_j}} )\propto \big| (M_\nu)_{ij} \big|^2$. In our case it is $M_N$ that gets probed, together with the $U_e$ matrix.

Using the formulas  \eqref{Ulformula}  and  \eqref{HD-matrix}, one obtains, to the leading order in $s_a t_{2 \beta}$
   \begin{equation}\label{doublychargeddecay}
  \frac{\Gamma_{\delta_L^{--}\rightarrow e_{L_i} e_{L_j}}}{\Gamma_{\delta_{R}\rightarrow e_{R_i} e_{R_j}}}
  \simeq \frac{m_{\delta_L^{--}} }{m_{\delta_R^{--} } } \!\bigg[1 
 \!+\!2 s_at_{2\beta} \, \text{Im} \frac{     \big( \mathcal{H}_D  M_N   +    M_N \mathcal{H}_D^T  \big)_{ij}    }{
  (M_N)_{ij}   }    \bigg]  
\end{equation}
This expression is of great importance, since it directly and manifestly probes, through the asymmetry or left and right doubly charged leptonic decays, the Hermitian part $M_D^H$ of the Dirac mass matrix. It is straightforward to obtain higher orders of the above expression.  

It cannot be overstressed: it is not just $M_D$, as usually assumed, that enters into the seesaw formula. The matrix $U_e$ also plays an essential role and there is a deep connection between these quantities which  can in principle be probed through the above doubly charged scalar decays. 

Of course, in the Hermitian limit $s_at_{2\beta} \simeq0 $, one simply ends up with a LR symmetric prediction of the same LH and RH doubly charged scalar decay rates.
 \vspace{0.3cm}
 \\
{ $\bullet$ Decays of the singly charged scalars $\delta_L^+$.}  
  \vspace{0.1cm}
  \\
The decay $\delta_L^-  \to e^-_L N$ which proceeds through the $\nu - N$ mixing $\Theta$ is of similar importance as the $N \to \ell_L  W$ decay since they both proceed through $M_D$. The $\delta^+_L$ interaction is given in \eqref{Ly},
and it is easily shown to lead
 to the decay rate
\begin{equation}
\Gamma_{\delta_L^- \to e_{Li} N_j} \propto \frac{m_{\delta_L^-}}{M_{W_R}^2} \big|(U_e^{\dagger}M_NU_e^{*} M_D^T M_N^{-1} V_R)_{ij} \big|^2
\end{equation}
which could in principle probe $M_D$ if $M_N$ was to be determined. It is a rather complicated expression in general and obviously hard to verify. It is illustrative to see what happens in the $P$ conserving situation and the same LH and RH mixing matrices $V_L = V_R$.  Using \eqref{MD-VL=VR} the above decay rate obtains a simple flavour structure $ \big| \big(V_L m_N \sqrt {v_L/v_R- m_{\nu}/m_N}\big)_{ij} \big|^2$.
%
%
%
 \vspace{0.3cm}
 \\
{$\bullet$ Decays of the neutral scalars $\delta_{L,R}^0$ }
\vspace{0.1cm}
\\
The decays $\delta_{L,R}^0  \to \nu N$ also proceed through  $\Theta$. 
The relevant interactions are given in \eqref{Ly}, 
which then gives the decay rate for $\delta_R^0$
\begin{equation}
\Gamma_{\delta_R^0\to \nu_{i} N_j}\propto \frac{m_{\delta_R^0}}{M_{W_R}^2} \big|\big(V_L ^\dagger  M_D^\dagger  V_R\big)_{ij} \big|^2
\end{equation}
It has the same flavor dependence as the $N \to h \nu$ decay. This is to be expected since $h$ and $\delta_R^0$ mix in general and the above decay can also proceed through this  mixing and the direct decay in~\eqref{N2Wdecay}. Since the mixing is proportional to the ratio of $W_L/W_R$ masses, the final result is naturally of the same order of magnitude. 

Similarly, one has for the left-handed $\delta_L^0$ decay  
\begin{equation}
\Gamma_{\delta_L^0\to \nu_{i} N_j} \propto \frac{m_{\delta_L^0}}{M_{W_R}^2} \big| \big(V_L ^\dagger U_e^{\dagger} M_N U_e^* M_D^T M_N^{-1} V_R\big)_{ij} \big|^2
\end{equation}
As expected, this decay has a rather different form due to the impact of $U_e$ and is quite messy, not easy to probe. It becomes illustrative in the $V_L=V_R$ limit, since by using \eqref{MD-VL=VR} the complicated flavour structure above becomes diagonal
$\big| \big(m_N \sqrt {v_L/v_R- m_{\nu}/m_N}\big)_{ij} \big|^2$.
 \vspace{0.3cm}
\\
{$\bullet$ Decay of the heavy charged scalars $H^{+}$}
\vspace{0.1cm}
\\
The relevant decays here are  $H^{-} \to e_L  N$  and $H^{-} \to e_R \nu$; they can be easily computed from \eqref{Phi-int} to be  
\begin{equation}
\Gamma_{H^- \to e_{Li}\, N_j} \propto \frac{m_H}{M_{W_L}^2} \big| [(m_e +e^{-ia} s_{2\beta} M_D^\dagger) V_R]_{ij} \big|^2
\end{equation}
and 
\begin{equation}
\Gamma_{H^- \to e_{Ri}\, \nu_j} \propto \frac{m_H}{M_{W_L}^2} \big| \big[ V_L^\dagger \big(M_D^\dagger +e^{ia} s_{2\beta} m_e\big) \big]_{ij} \big|^2
\end{equation}

It is interesting to check what happens on the Hermitian limit $s_at_{2\beta}\simeq0$. The result depends how $s_at_{2\beta}$ vanishes: if $a\simeq0$ nothing new happens, unlike the opposite possibility $\beta\simeq0$. In this case the decay $H^{-} \to e_L  N$ measures directly $V_R$, while the $H^{-} \to e_R \nu$ decay probes $M_D$, which, as we know from \ref{unbrokenP}, gets predicted from $M_\nu$ and $M_N$.
\vspace{0.3cm}
 \\
{  $\bullet$ Decays of the heavy neutral scalar $H^0$}
\vspace{0.1cm}
\\
The relevant interactions can be easily determined from \eqref{Phi-int}.
The decay $H^{0} \to e \bar e$ is then given by
\begin{align}\begin{split}\label{Htoeebar}
\Gamma_{H^0\rightarrow e_i\bar{e}_j}
\propto \frac{m_H}{M_{W_L}^2}  \big|(M_D^\dagger + e^{i a} s_{2\beta}m_{e})_{ij}\big|^2
\end{split}
\end{align}
 The decay $H^{0} \to \nu N$ 
 is probably of secondary importance, but nonetheless we give the relevant decay rate  
\begin{align}\begin{split}
\Gamma_{H^0\rightarrow \nu_i N_j}
\propto \frac{m_H}{M_{W_L}^2}  \big|\big[V_L^\dagger \big(m_e +e^{-ia} s_{2\beta} M_D^\dagger\big)  V_R\big]_{ij}\big|^2
\end{split}
\end{align}
   As before, the limit $\beta\simeq0$ simplifies matters. The $H^{0} \to e \bar e$ decay would then directly probe $M_D$,while the $H^{0} \to \nu N$ decay depends on the product of $V_L$ and $V_R$, which makes it harder to disentagle. We should stress once again that $\beta\simeq0$ requires heavy $W_R$, beyond the LHC reach. Furthermore, the same flavour constraints indicate~\cite{Bertolini:2014sua} that light $W_R$ in the LHC energy reach basically fixes (modulo uncertainties) $s_a t_{2\beta} \simeq 0.02$.
This in return reduces substantially the freedom in the quark RH mixing and $M_D$. 
      
In short, there are a plethora of high energy physical processes that probe $M_D$, on top of measuring $M_N$ as in the original KS case. If these states were to be accessible at the LHC or the next hadron collider, one would have a clear shot of verifying or disproving the MLRSM. Here we have given the list and the associated decay rates for these processes, their phenomenology shall be dealt with in detail in a forthcoming publication.

\subsection{The limits on particle masses} \label{masslimits}

   Before we turn to the conclusions it may be useful to the reader to have an idea of the limits on the masses of the relevant gauge boson and scalar particles.
   
   \vspace{0.3cm}
 
{$\bullet$ Limits on the masses of heavy charged and neutral gauge bosons $W_R$ and $Z_R$.}
   \vspace{0.2cm}

   The best limit on the $W_R$ mass comes from the di-jet final state and is thus independent of the detailed properties of the RH neutrinos, and amounts to $M_{W_R} \geq 4$\,TeV~\cite{Aad:2019hjw}. The limit from the KS process depends on the mass of the RH neutrinos and it varies from $3.5$ to $5$ TeV for $m_N$ in the range from $0.1$ to $1.8$ TeV~\cite{Aaboud:2019wfg}.
   
   The limit on $Z_R$ mass is  $M_{Z_R}\!\geq\!4.6$\,TeV\,\cite{Aaboud:2019wfg}. We should stress that the MLRSM prediction $M_{Z_R} \simeq 1.7 M_{W_R}$ implies that the LHC cannot see the $Z_R$, and thus, if it was to be seen, it would automatically invalidate the MLRSM. 
   
     \vspace{0.3cm}
 
 {  $\bullet$ Limit on the mass of the heavy doublet $H$ from the bi-doublet $\Phi$.}
 
 \vspace{0.2cm}
 
  As discussed in section \ref{section:mlrsm}, since the couplings of the $H$, given in \eqref{Phi-int}, are determined by the structure of the bi-doublet, the flavor conservation in neutral currents sets a stringent limit $m_H \geq 20$ TeV~\cite{OtherLR, Bertolini:2014sua}. This raises the question of the perturbativity of the theory for a $W_R$ accessible at the LHC, i.e., for $M_{W_R} \lesssim 8 $\,TeV, since $H$ and $W_R$ get the mass at the same large stage of symmetry breaking, and thus a particular scalar coupling must be large enough to ensure the heaviness of $H$. 
 This implies stringent limits on other scalar masses to be discussed below.
   For a heavy $W_R$, with $ M_{W_R} \gtrsim 20$\,TeV, so that the $H$ mass does not require a large coupling, these limits go away and the theory is of course highly perturbative.

    \vspace{0.2cm}
 
 { $\bullet$ Limits on the masses of the heavy scalar triplet $\Delta_L$.}
 
 \vspace{0.2cm}
 
 The best direct limit is on the mass of the doubly charged component and is roughly $m_{\delta_L ^{++}} \geq 400$\,GeV, but is flavor dependent~\cite{Maiezza:2016bzp}. Since the T parameter is sensitive to the mass splittings in the $\Delta_L$ multiplet, one obtains better limit from the high precision T constraints, see Fig. 6 in~\cite{Maiezza:2016bzp}. For a relatively light $W_R$ accessible at the LHC, this implies a limit on the (basically degenerate) multiplet mass on the order of TeV. 
 
   As discussed above, the heaviness of the second doublet $H$ brings the issue of perturbativity and a much stronger limit emerges.
   By asking that the perturbative cutoff be not below 10 $M_{W_R}$, for $M_{W_R} = 6 $\,TeV one gets $m_{\Delta_L} \geq 9 $\,TeV~\cite{Maiezza:2016ybz}, far from the LHC reach. 
For a $W_R$ at the LHC energies, seeing the $\Delta_L$ multiplet would invalidate the MLRSM. We should stress though that this is not a generic prediction of the theory. For a heavy $W_R$ above say 20 TeV the only limit that remains is the direct one.

   \vspace{0.3cm}
 
 {  $\bullet$ Limits on the masses of the $\delta_R^{++}$ and $\delta_R^0$.}
 
 \vspace{0.2cm}
 
  The direct limit on the mass of $\delta_R^{++}$ is similar to  the one of its left-handed counterpart, of roughly 400 GeV,  also flavor dependent~\cite{Maiezza:2016bzp}. However, similar to the situation regarding $\Delta_L$, the perturbativity bound becomes huge for a $W_R$ at the LHC and in this case $m_{\delta_R^{++}} \geq 12$\,TeV~\cite{Maiezza:2016ybz}. Once again, the limit disappears for heavy $W_R$ above 20 TeV. 
  
  On the other hand, $\delta_R^0$ can be arbitrarily light. One could hope for a cosmological limit from the stability of the scalar potential, but we may be living in a meta-stable vacuum.

\section{Summary and outlook}
\label{section:outlook}

The origin and nature of neutrino mass is arguably one of the central issues in the quest for the theory beyond the Standard Model. Over the years, the seesaw mechanism has emerged as the main scenario behind the smallness of neutrino mass, but by itself falls short of providing a full-fledged theory. First of all, the SM seesaw cannot be disentangled, and furthermore, the heavy RH neutrinos can be produced at the hadron colliders, such as the LHC, only 
through the Dirac mass terms. This obviously forces
 the Dirac mass terms to be incomparably larger than their natural seesaw values. 
 
The situation changes dramatically in the context of the LR symmetric theory, the same one that led originally to non-vanishing neutrino mass and to the seesaw mechanism itself. The SM singlet RH neutrinos can be easily and naturally produced at the hadron colliders due to their gauge interactions with the $W_R$ boson, the RH counterpart of the SM $W$ boson. Equally important, the Dirac mass terms stop being plagued by ambiguities allowing for a verifiable seesaw based on the Higgs mechanism as the origin of neutrino mass. This is manifest in the case of LR symmetry being charge conjugation, but it turns out to be quite subtle in the case of parity.

Whereas for $\mathcal{C}$ it is straightforward to compute $M_D$ as a function of $M_\nu$ and $M_N$ in the Minimal LR Symmetric Model, the same was never achieved for the $\mathcal{P}$ case, at least not in full generality. Instead, we recently suggested an alternative approach using the exciting LNV violating decays of doubly charged scalars into the pair of the same sign charged leptons as a probe of $M_D$. Moreover we managed to solve analytically the parity conserving case $M_D = M_D^\dagger$, and to show that, when parity gets broken, only the Hermitian part $M_D^H$ is independent, simplifying thus considerably the experimental verification of the program of disentangling the seesaw.

In this longer follow-up of our  original work, we have filled in the missing gaps in obtaining our results. We have given a pedagogical expose of the Hermitian case, illustrating it with a few illuminating examples, and providing the relevant derivations. We have also provided a list of  relevant decays that depend on $M_D$, and can thus serve to probe the seesaw origin of neutrino mass. A number of these processes can test directly the Majorana nature of $N$'s and  are  complementary to the neutrinoless double beta decay. This is true in particular of the so-called KS process of the direct production of $N$'s, but it applies also to the decays of doubly and singly charged scalars from the heavy Higgs triplets $\Delta_{L,R}$. 

In principle also the neutral scalar decays can play the same role, especially the decay of $\delta_R^0$ into the pairs of $N$'s, and possibly even through the same decays of the SM Higgs boson, in the case of appreciable $h - \delta_R^0$ mixing (for a recent phenomenological study, see~\cite{Maiezza:2015lza}). All these particles can also decay into  final states that probe $M_D$, as discussed in Section V. Moreover, the decays of the heavy scalar doublet $H$ from the bi-doublet $\Phi$ also probe directly $M_D$, but due to the large limit on its mass are out of LHC reach.

Before closing it is worthwhile to compare the MLRSM with other gauge theories of neutrino mass.
  As we argued from the beginning it is the LR symmetry that leads to the existence of RH neutrinos $N$ which then lead to the seesaw and the rest of the story told here. In turn the global $B-L$ symmetry of the SM is gauged automatically, and the cancellation of new induced anomalies provides an additional raison d'etre for $N$ besides the LR symmetry. What happens then in the
 more modest approach where one gauges just the  $G_{SM} \times U(1)_{B-L}$ subgroup of the MLRSM? First of all, $N$ can be produced pairwise through the new neutral gauge boson $Z'$, an improvement over the simple SM seesaw. Moreover, if one were to observe  the $N \to W^+ e_L$ process, this would require the $\nu-N$ mixing, and in a renormalizable theory that mixing would require the Yukawa Dirac coupling with the SM Higgs. Thus one could argue that the above decay would probe the Higgs origin of neutrino mass, just as in the MLRSM. However, there is a profound difference. In the $U(1)_{B-L}$ gauged model $M_D$ is arbitrary as in the SM seesaw, i.e. one has the situation described in \eqref{MD-I}. It is precisely here where the MLRSM stands out by having $M_D$ structurally predicted from the knowledge of $M_\nu$ and $M_N$, which is per se  a great motivation for studying MLRSM, besides the desire to understand the origin  of the breaking of parity in weak interactions. 
 
   One could also be more ambitious and embed the MLRSM in the $SO(10)$ grand unified theory which besides providing the $N$'s and the seesaw, also links quark and charged lepton masses. This would be extremely exciting if the scale of LR symmetry breaking was accessible to present day or near future colliders. Unfortunately, in the minimal predictive version of the $SO(10)$ theory the scale of breaking of $SU(2)_R$ is predicted to be huge, above $10^{10} $GeV or so, and thus hopelessly out of reach. The beauty of predicting $M_D$ or using $M_D$ to determine $M_N$ ends up being more a question of aesthetics than of the directly verifiable physics.
   
   Another natural theory of neutrino mass is the Minimal Supersymmetric Standard Model without the artificial assumption of R-parity conservation. It provides the mixture of both the seesaw and the radiative origin of neutrino masses, and it offers the connection between the neutrinoless double beta decay and lepton number violation at hadronic colliders \cite{Allanach:2009iv}. Unfortunately the proliferation of unknown parameters, such as the masses and couplings of the superpartners, impedes the predictivity.
 
    All in all,  our work offers yet another conclusive evidence in favor of the MLRSM standing out by being a self-contained theory of neutrino mass, in complete analogy with the SM as a theory of the origin of charged fermion masses.

\subsection*{Acknowledgments}

  We acknowledge the collaboration with Miha Nemev\v{s}ek at the initial stages of this work.  We thank Alessio Maiezza, Alejandra Melfo, Fabrizio Nesti and Juan Carlos Vasquez for useful comments and discussions, and a careful reading of the manuscript. 
GS is deeply grateful to Biba and Miki center for healthy and happy living in the village of Vrboska, island of Hvar, for a great hospitality extended to him during the last stages of this work. 
He also acknowledges the hospitality of Gira (Vrboska) and Karaka (Split) establishments.

\end{document}